\documentclass[bjps]{imsart}

\RequirePackage[OT1]{fontenc}
\usepackage{color}
\usepackage{amsmath}
\usepackage{amssymb}
\usepackage{graphicx}
\usepackage{float}
\usepackage{multirow}
\usepackage[authoryear]{natbib}
\usepackage[T1]{fontenc}
\usepackage{amsbsy,amsfonts,amscd,amsthm,mathrsfs,epsfig,anysize}
\usepackage{url}
\RequirePackage{hyperref}

\arxiv{arXiv:0000.0000}

\startlocaldefs
\numberwithin{equation}{section}
\theoremstyle{plain}

\newcommand{\summ}[2]{\sum _{#1}^{#2}} 
\newcommand{\prodd}[2]{\prod _{#1}^{#2}} 
\newcommand{\bs}{\boldsymbol} 

\endlocaldefs

\begin{document}

\begin{frontmatter}
\title{Inference on Dynamic Models for non-Gaussian Random Fields using INLA: A Homicide Rate Anaysis of Brazilian Cities}
\runtitle{Dynamic Models with INLA}

\begin{aug}
\author{\fnms{Cortes,} \snm{R. X.}\thanksref{a}\ead[label=e1]{renanxcortes@gmail.com}},
\author{\fnms{Martins,} \snm{T. G.}\thanksref{b}\ead[label=e2]{thigm85@gmail.com}},
\author{\fnms{Prates,} \snm{M. O.}\thanksref{a}\ead[label=e3]{marcosop@gmail.com}}
\and
\author{\fnms{Silva} \snm{B. A.}\thanksref{c}%
\ead[label=e4]{braulio.fas@gmail.com}%
\ead[label=u1,url]{http://www.foo.com}}

\runauthor{Cortes, R. X., et al.}

\affiliation[a]{Department of Statistics, Universidade Federal de Minas Gerais}
\affiliation[b]{Department of Mathematical Sciences, Norwegian University of Science and Technology}
\affiliation[c]{Department of Sociology, Universidade Federal de Minas Gerais}

\address{Department of Statistics\\
Universidade Federal de Minas Gerais\\
Presidente Ant\^{o}nio Carlos, 6627, Belo Horizonte, Brazil\\
\printead{e1,e3}}

\address{Department of Mathematical Sciences\\
Norwegian University of Science and Technology\\
H{\o}gskoleringen 1, 7491, Trondheim, Norway\\
\printead{e2}}

\address{Department of Sociology\\
Universidade Federal de Minas Gerais\\
Presidente Ant\^{o}nio Carlos, 6627, Belo Horizonte, Brazil\\
\printead{e4}}

\end{aug}

\begin{abstract}
Robust time series analysis is an important subject in statistical modeling. 
Models based on Gaussian distribution are sensitive to outliers, which may 
imply in a significant degradation in estimation performance as well as in
prediction accuracy. State-space models, also referred as Dynamic Models, 
is a very useful way to describe the evolution
of a time series variable through a structured latent evolution system.
Integrated Nested Laplace Approximation (INLA) is a recent approach
proposed to perform fast approximate Bayesian inference in Latent Gaussian Models
which naturally comprises Dynamic Models. We present how to perform fast 
and accurate non-Gaussian dynamic modeling with INLA and show how these 
models can provide a more robust time series analysis when compared with 
standard dynamic models based on Gaussian distributions. We formalize the 
framework used to fit complex non-Gaussian space-state models using the R 
package INLA and illustrate our approach in both a simulation study and a 
Brazilian homicide rate dataset.
\end{abstract}




\end{frontmatter}

\section{Introduction}

\label{sec:Introduction}

Robust estimation of time series analysis is an important and challenging
field of statistical application from either frequentist \citep{bustos1986,denby1979}
or Bayesian perspectives \citep{west1981}. Such models are preferred
when the dataset under study is affected by structural or abrupt
changes. Dynamic Linear Models (DLM) and Generalized Dynamic Linear Models (DGLM), 
also referred as state-space models, are a broad class of
parametric models that generalizes regression and time series models
with time varying parameters, where both the parameter variation and
the observed data are described in an evolution structured way \citep{migon2005}.
Dynamic models are composed by an observational equation and one or
more system equations in which the error terms are usually chosen
to follow a Gaussian distribution. However, it is well known that
the Gaussian distribution is very sensitive to outliers, which may
produce degradation in the estimation performance \citep{fox1972}.
Therefore, one might be interested in building a more flexible model
based on heavy-tailed distributions rather than the usual Gaussian.
Such models fall into the class of \textit{non-Gaussian dynamic models} 
(see, \citet{kitagawa1987non,durbin2000} for a detailed
description as well as applications of this class of models).

Integrated Nested Laplace Approximation (INLA) is an approach proposed
by \citet{rue2009approximate} to perform approximate fully Bayesian
inference in the class of latent Gaussian models (LGMs). LGMs is a
broad class and include many of the standard models currently in use
by the applied community, e.g., stochastic volatility, disease mapping,
log-Gaussian Cox process and generalized linear models. As opposed
to the simulation-based methods, like Markov Chain Monte Carlo (MCMC),
INLA performs approximate inference using a series of deterministic
approximations that take advantage of the LGM structure to provide
fast and accurate approximations. Moreover, it avoids known problems with
commonly used simulation-based methods, e.g., difficulty in diagnosing
convergence, additive Monte Carlo errors, and high demand in terms
of computational time. Even for dynamic models within the class of LGMs, it was not possible to fit most of them using
the available tools in the \texttt{INLA} package for \texttt{R}, hereafter
denoted as \texttt{R-INLA}. \citet{ruiz2011direct} presented a general framework
which enabled users to use \texttt{R-INLA} to perform fully Bayesian
inference for a variety of state-space models. However, their approach does 
not include the class of\textit{ non-Gaussian dynamic models} where the errors 
of the system equations have a non-Gaussian distribution as, for example, the 
heavy-tailed distributions.

One of the key assumptions of the INLA approach is that the latent
field follows a Gaussian distribution. However, \citet{martins2014extendingScan}
have shown a way to extend INLA to cases where some independent components
of the latent field have a non-Gaussian distribution. Their approach
transfer the non-Gaussianity of the latent field to the likelihood
function and it has shown to produce satisfactory results as long
as this distribution is not far from Gaussian. Distributions that add flexibility around a Gaussian as \textit{near-Gaussian} distributions are referred as being, for example, unimodal and symmetric.

The contribution of this paper is three folded: 1) Extend INLA for non-Gaussian 
latent models with dependent structure, specifically for non-Gaussian DLMs; 2) 
Present in a simple manner how to use \texttt{R-INLA} to perform non-Gaussian DLMs
modelling.
To accomplish these issues, we introduce a reparametrization of the 
non-Gaussian DLM and combine with the computational framework provided by 
\texttt{R-INLA} to introduce how to model dependent non-Gaussian latent field 
in the \texttt{R-INLA} setup;
3) We analyze the Brazilian homicides rates using a robust approach. 
The analysis indicates that, in most of our application scenarios, 
the robust method outperforms the traditional Gaussian approach.

The paper is organized as following: Section \ref{sec:methodology} introduces the 
methodology of our approach, presenting how to perform fast Bayesian inference using 
\texttt{R-INLA} for non-Gaussian DLMs. Section \ref{sec:Simulations} presents our 
simulation study to compare two competitor models using some quality measures. In Section 
\ref{sec:data_application} we present the study over the Brazilian Homicide data, 
explaining our findings. Finally, in Section 5 we discuss some final remarks and future research.

\section{Methodology}\label{sec:methodology}

This Section will describe our approach to handle non-Gaussian DLM within 
\texttt{R-INLA}. Although valid for DGLM, we have chosen
to illustrate our extension using a DLM to facilitate the presentation.
To apply the extension for DGLM, a simple change in the Gaussian likelihood
is necessary, which is a trivial modification under \texttt{R-INLA}. 
Section \ref{sec:Models} will define a general DLM of interest, and show 
that it fits the class of LGM only if the error terms of the system 
equations are Gaussian distributed. Section \ref{sec:INLAreview} will 
review the INLA methodology, including the recent extension 
that allows INLA to be applied to models where some components of the 
latent field have non-Gaussian distribution. 
Section \ref{sec:INLAforDLMs} gives an overview of a generic 
approach to fit dynamic models using \texttt{R-INLA} through an augmented 
model structure. Finally, 
Section \ref{sec:Contribution} extend the 
approaches presented in Sections \ref{sec:INLAreview} and 
\ref{sec:INLAforDLMs} and show how this extension can be exploited
to fit non-Gaussian DLM within \texttt{R-INLA}.

\subsection{Models}\label{sec:Models}

INLA approach performs approximate Bayesian inference in latent Gaussian models where the first stage is formed by the likelihood function
with conditional independence properties given the latent field $\bs{x}$
and possible hyperparameters $\bs{\theta}_{1}$, where each data point
$\{y_{t},\ t = 1, ..., n_d\}$ is connected to one element in the latent
field. In this context, the latent field $\bs{x}$ is formed by linear
predictors, random and fixed effects, depending on the model formulation. 
Assuming that the elements of the latent field connected
to the data points, i.e., the linear predictors $\{\eta_t,\ t=1,...,n_d\}$, 
are positioned on the first $n_{d}$ elements of $\bs{x}$, we have \begin{list}{\labelitemi}{\leftmargin=5em}

\item {\textbf{Stage 1.}} $\bs{y}|\bs{x},\bs{\theta}_{1}\sim\pi(\bs{y}|\bs{x},\bs{\theta}_{1})=\prodd{t=1}{n_{d}}\pi(y_{t}|x_{t},\bs{\theta}_{1})$.
\end{list} The conditional distribution of the $\bs{x}$ given some possible hyperparameters $\bs{\theta}_{2}$ forms the second
stage of the model and has a joint Gaussian distribution, 
\begin{list}{\labelitemi}{\leftmargin=5em}
\item {\textbf{Stage 2.}} $\bs{x}|\bs{\theta}_{2}\sim\pi(\bs{x}|\bs{\theta}_{2})=\mathcal{N}(\bs{x};\bs{\mu}(\bs{\theta}_{2}),\bs{Q}^{-1}(\bs{\theta}_{2}))$,
\end{list} where $\mathcal{N}(\cdot;\bs{\mu},\bs{Q}^{-1})$ denotes
a multivariate Gaussian distribution with mean vector $\bs{\mu}$
and a precision matrix $\bs{Q}$. In most applications, the latent
Gaussian field have conditional independence properties, which translates
into a sparse precision matrix $\bs{Q}(\bs{\theta}_{2})$, which is
of extreme importance for the numerical algorithms used by INLA. The latent field $\bs{x}$ may have additional linear constraints
of the form $\bs{A}\bs{x}=\bs{e}$ for an $q\times n_d$ matrix $\bs{A}$
of rank $q$, where $q$ is the number of constraints and $n_d$ the
size of the latent field. The hierarchical model is then completed
with an appropriate prior distribution for the $m$-dimensional hyperparameter of
the model $\bs{\theta}=(\bs{\theta}_{1},\bs{\theta}_{2})$ \begin{list}{\labelitemi}{\leftmargin=5em}

\item {\textbf{Stage 3.}} $\bs{\theta}\sim\pi(\bs{\theta}).$ \end{list}

The structure of a non-Gaussian DLM is composed by an observation
equation describing the relationship between the observations 
$\bs{y}$ $\{y_t;\ t = 1, ..., n_d\}$, which are connected to a linear combination 
of the state parameters $\bs{a}$ $\{a_t;\ t = 1, ..., n_d\}$, and a system of 
equations describing the evolution of $\bs{a}$, for example: 
\begin{align*}
y_{t}=a_{t}+v_{t}, & \qquad v_{t}\sim N(0,\theta_{1})\\
a_{t}=a_{t-1}+w_{t}, & \qquad w_{t}\sim \pi(\cdot)
\end{align*}
in which $a_{t}$ is the state vector at time
t, $\theta_{1}$ is the Gaussian variance of $v_{t}$ and the noises $w_{t}$ could
follow a non-Gaussian distribution. We emphasize that the structure described above could be more flexible allowing any linear combination and addition of covariates. Furthermore, this is an extension over the traditional DLM where we now can have a non-Gaussian distribution for the noise $w_{t}$ in the system equation .

If $w_{t}$ is assumed to be Gaussian,
this structure falls naturally into the class of LGMs (see Section 
\ref{sec:INLAforDLMs}). To help understand the INLA review of Section
\ref{sec:INLAreview} we can rewrite a LGM using a hierarchical structure
with three stages. To elucidate the understanding of notation in our examples, we highlight that state vector $\bs{a}$ does not necessarily corresponds to the latent field $\bs{x}$. Since our approach lies in a augmented likelihood function, the dimension of the latent field $\bs{x}$ is larger than the dimension of $\bs{y}$ and $\bs{a}$.

However, if a Gaussian distribution is not assumed for $w_{t}$,
it is no longer possible to write the model as a hierarchical structure 
with the Gaussian assumption in the second stage. To accommodate the non-Gaussian
DLM it is necessary to expand the class of LGMs defined
early to allow that nodes of the latent field have non-Gaussian
distributions. We then rewrite stage 2 of the hierarchical model
as \begin{list}{\labelitemi}{\leftmargin=5em}
\item {\textbf{Stage} $\bs{2^{new}}$.} $\underbrace{(\bs{x}_{G},\bs{x}_{NG})}_{\bs{x}}|\bs{\theta}_{2}\sim\pi(\bs{x}|\bs{\theta}_{2})=\mathcal{N}(\bs{x}_{G};\bs{0},\bs{Q}^{-1}(\bs{\theta}_{2}))\times\prod_{t}\pi(\bs{x}_{NG_t}|\bs{\theta}_{2})$,
\end{list} where $\bs{x}_{G}$ and $\bs{x}_{NG}$ represent the Gaussian
and independent non-Gaussian terms of the latent field, respectively. As a result,
the distribution of the latent field is not Gaussian, which
precludes the use of INLA to fit this class of models.


Section \ref{sec:INLAreview} summarizes how to perform inference, 
within the \texttt{R-INLA} framework, on models where the non-Gaussian components of 
the latent field belong to the class of near-Gaussian distributions. Later, 
in Section \ref{sec:Contribution} we introduce how to perform inference 
when the non-Gaussian components have a dependent structure, specifically
belong to the class of non-Gaussian DLMs.

\subsection{INLA review}\label{sec:INLAreview}

Using the hierarchical representation of LGMs given in Section \ref{sec:Models}
we have that the joint posterior distribution of the unknowns is
\begin{align*}
\pi(\bs{x},\bs{\theta}|\bs{y}) & \propto\pi(\bs{\theta})\pi(\bs{x}|\bs{\theta})\prodd{t=1}{n_{d}}\pi(y_{t}|x_{t},\bs{\theta})\\
 & \propto\pi(\bs{\theta})|\bs{Q}(\bs{\theta})|^{n/2}\exp\bigg[-\frac{1}{2}\bs{x}^{T}\bs{Q}(\bs{\theta})\bs{x}+\summ{t=1}{n_{d}}\log\{\pi(y_{t}|x_{t},\bs{\theta})\}\bigg].
\end{align*}
The approximated posterior marginals of interest $\tilde{\pi}(x_{t}|\bs{y})$,
$t=1,..,n_{d}$ and $\tilde{\pi}(\theta_{j}|\bs{y})$, $j=1,...,m$ returned
by INLA have the following form 
\begin{align}
\tilde{\pi}(x_{t}|\bs{y}) & =\sum_{u}\tilde{\pi}(x_{t}|\bs{\theta}^{(u)},\bs{y})\tilde{\pi}(\bs{\theta}^{(u)}|\bs{y})\ \Delta\bs{\theta}^{(u)}\label{eq:INLAximarg}\\
\tilde{\pi}(\theta_{s}|\bs{y}) & =\int\tilde{\pi}(\bs{\theta}|\bs{y})d\bs{\theta}_{-s}
\label{eq:INLAthetajmargcont}
\end{align}
where $\{\tilde{\pi}(\bs{\theta}^{(u)}|\bs{y})\}$ are the density
values computed during a grid exploration on $\tilde{\pi}(\bs{\theta}|\bs{y})$,
for given approximations of $\pi(x_{t}|\bs{\theta},\bs{y})$ and $\pi(\bs{\theta}|\bs{y})$.

Looking at Eqs.~\eqref{eq:INLAximarg}-\eqref{eq:INLAthetajmargcont}
we can see that the method can be divided into three main tasks. First, propose an 
approximation $\tilde{\pi}(\bs{\theta}|\bs{y})$ to the joint posterior 
of the hyperparameters $\pi(\bs{\theta}|\bs{y})$, second propose an 
approximation $\tilde{\pi}(x_{t}|\bs{\theta},\bs{y})$
to the marginals of the conditional distribution of the latent field
given the data and the hyperparameters $\pi(x_{t}|\bs{\theta},\bs{y})$
and last explore $\tilde{\pi}(\bs{\theta}|\bs{y})$ on a grid and
use it to integrate out $\bs{\theta}$ in Eq.~\eqref{eq:INLAximarg}
and $\bs{\theta}_{-j}$ in Eq.~\eqref{eq:INLAthetajmargcont}.

The approximation used for the joint posterior of the hyperparameters
$\pi(\bs{\theta}|\bs{y})$ is 
\begin{equation}
\tilde{\pi}(\bs{\theta}|\bs{y})\propto\frac{\pi(\bs{x},\bs{\theta},\bs{y})}{\pi_{G}(\bs{x}|\bs{\theta},\bs{y})}\bigg|_{\bs{x}=\bs{x}*(\bs{\theta})}\label{eq:lapthetay}
\end{equation}
where $\pi_{G}(\bs{x}|\bs{\theta},\bs{y})$ is a Gaussian approximation
to the full conditional of $\bs{x}$, and $\bs{x}^{*}(\bs{\theta})$
is the mode of the full conditional for $\bs{x}$, for a given $\bs{\theta}$.
The full conditional of the latent field when dealing with LGMs is given by 
\begin{equation}
\pi(\bs{x}|\bs{\theta},\bs{y})\propto\exp\bigg\{-\frac{1}{2}\bs{x}^{T}\bs{Q}(\bs{\theta})\bs{x}+\sum_{t\in \mathcal{T}}g_{t}(x_{t})\bigg\},\label{eq:fullcondx}
\end{equation}
where $\mathcal{T}$ is an index set and $g_{t}(x_{t})=\log\pi(y_{t}|x_{t},\bs{\theta_{1}})$. The
Gaussian approximation used by INLA is obtained by matching the modal
configuration and the curvature at the mode. The good performance
of INLA is highly dependent on the appropriateness of the Gaussian approximation
in Eq.~\eqref{eq:fullcondx} and this turns out to be the case when
dealing with LGMs because the Gaussian prior assigned to the latent
field has a non-negligeable effect on the full conditional, specially
in terms of shape and correlations. Besides, the likelihood function
is usually well behaved and not very informative on $\bs{x}$. It
is very important to note that Eq.~\eqref{eq:lapthetay} is equivalent 
to the Laplace approximation of a marginal posterior 
distribution \citep{tierney1986accurate}, and it is exact if $\pi(\bs{x}|\bs{y},\bs{\theta})$ is Gaussian, 
in which case INLA gives exact results up to small
integration error due to the numerical integration of Eq.~\eqref{eq:INLAximarg}
and Eq.~\eqref{eq:INLAthetajmargcont}.

For approximating $\pi(x_{t}|\bs{\theta},\bs{y})$, three options are available 
in \texttt{R-INLA}. The so called Laplace, Simplified Laplace and Gaussian which 
are ordered in terms of accuracy. We refer to \citet{rue2009approximate} for a 
detailed description of these approximations and \citet{martins2012bayesian} 
on how to compute Eq.~\eqref{eq:INLAthetajmargcont} efficiently.

\citet{martins2014extendingScan} have demonstrated how INLA can be used
to perform inference in latent models where some independent components
of the latent field have a non-Gaussian distribution, in which case
the latent field is no longer Gaussian. Their approach approximates
the distribution of the non-Gaussian components $\pi(\bs{x}_{NG}|\bs{\theta}_{2})$
by a Gaussian distribution $\pi_{G}(\bs{x}_{NG}|\bs{\theta}_{2})$
and corrects this approximation with the correction term 
\begin{equation*}
CT=\pi(\bs{x}_{NG}|\bs{\theta}_{2})/\pi_{G}(\bs{x}_{NG}|\bs{\theta}_{2})\label{eq:ext:CT}
\end{equation*}
in the likelihood. Taking into consideration the above approximation
and correction term we can rewrite our latent model with the following
hierarchical structure \begin{list}{\labelitemi}{\leftmargin=5em}
\item {\textbf{Stage 1.}} $\bs{z}|\bs{x},\bs{\theta}\sim\pi(\bs{z}|\bs{x},\bs{\theta})=\prodd{t=1}{n_{d}+k}\pi(z_{t}|x_{t},\bs{\theta})$,
where \end{list} 
\begin{equation*}
\pi(z_{t}|x_{t},\bs{\theta})=\Bigg\{\begin{array}{cc}
\pi(y_{t}|x_{t},\bs{\theta}_{1}) & \mbox{ for }1\leq t\leq n_{d}\\
\pi(x_{NG_t}|\bs{\theta}_{2})/\pi_{G}(x_{NG_t}|\bs{\theta}_{2}) & \mbox{ for }n_{d}<t\leq n_{d}+k
\end{array}\label{eq:fakelikelihood-1}
\end{equation*}
and $\bs{z}$ is an augmented response vector with $z_{t}=y_{t}$ if
$t\leq n_{d}$ and $z_{t}=0$ if $n_{d}<t\leq n_{d}+k$, where $k$
is the length of $\bs{x}_{NG}$. It is important to emphasize that
Stage 1 above is not the likelihood function, but expressing the model
using this form makes the practical definition of the non-Gaussian
latent model within the \texttt{R-INLA} framework easier to understand. 

The latent field has now a Gaussian approximation replacing the non-Gaussian
distribution of $\bs{x}_{NG}$, 
\begin{list}{\labelitemi}{\leftmargin=5em}
\item {\textbf{Stage 2.}} $\underbrace{(\bs{x}_{G},\bs{x}_{NG})}_{\bs{x}}|\bs{\theta}_{2}\sim\pi(\bs{x}|\bs{\theta}_{2})=\mathcal{N}(\bs{x}_{G};\bs{0},\bs{Q}^{-1}(\bs{\theta}_{2}))\times\pi_{G}(\bs{x}_{NG}|\bs{\theta}_{2})$,
\end{list} which means that $\pi(\bs{x}|\bs{\theta}_{2})$ is now
Gaussian distributed.

\citet{martins2014extendingScan} have shown that the main impact of this
strategy occurs in the Gaussian approximation to the full conditional
of the latent field that now takes the form 
\begin{equation}
\pi(\bs{x}|\bs{\theta},\bs{y})\propto\exp\bigg\{-\frac{1}{2}\bs{x}^{T}\bs{Q}(\bs{\theta})\bs{x}+\summ{t=1}{n_{d}}g_{t}(x_{t})+\summ{t=n_{d}+1}{n_{d}+k}h_{t}(x_{t})\bigg\},\label{eq:INLAext:fullcondx}
\end{equation}
where $g_{t}(x_{t})=\log\pi(y_{t}|x_{t},\bs{\theta})$ as before and
\[
h_{t}(x_{t})=\log CT_{t}=\log\pi(\bs{x}_{NG_t}|\bs{\theta}_{2})-\log\pi_{G}(\bs{x}_{NG_t}|\bs{\theta}_{2}).
\]
The key for a good accuracy of INLA depends
on the behavior of $h_{t}(x_{t})$ which is influenced by the distribution
$\pi(\bs{x}_{NG_t}|\bs{\theta}_{2})$ of the non-Gaussian components
and by the Gaussian approximation $\pi_{G}(\bs{x}_{NG_t}|\bs{\theta}_{2})$
to this non-Gaussian distribution. Also good results 
are obtained when $\pi_{G}(\bs{x}_{NG_t}|\bs{\theta}_{2})$ is chosen to be a
zero mean and low precision Gaussian distribution such that 
\[
\pi_{G}(\bs{x}_{NG_t}|\bs{\theta}_{2})\propto\text{constant}
\]
and $\pi(\bs{x}_{NG_t}|\bs{\theta}_{2})$ is not too far away from
a Gaussian, for which they coined the term \textit{near-Gaussian}
distributions. This means that the application of INLA within the 
context of non-Gaussian DLM will yield accurate results as long as 
these components are distributed according to a flexible distribution 
around the Gaussian, as in the Student's t case for example, which is unimodal and symmetric.

\subsection{\texttt{R-INLA} for DLM}\label{sec:INLAforDLMs}

In this section, we present a simple dynamic model to illustrate the framework 
to perform fast Bayesian inference within \texttt{R-INLA}. The INLA approach could 
be used to estimate any dynamic structure that could be written as a latent Gaussian 
model described in Section \ref{sec:Models}, however the approach presented here is 
motivated to overcome some limitations of \texttt{R-INLA}. Suppose as a Toy Example 
the following first order univariate dynamic linear model 
\begin{align}
y_{t} & = a_{t} + v_{t},\quad v_{t}\sim N(0,\theta_{1}),\ t=1,...,n_d\label{eq:fodlm:obseq}\\
a_{t} & = a_{t-1} + w_{t},\quad w_{t}\sim N(0,\theta_{2}),\ t=2,...,n_d.\label{eq:fodlm:sustemeq}
\end{align}
It is possible to fit the model given by Eqs.~\eqref{eq:fodlm:obseq} and
\eqref{eq:fodlm:sustemeq} using the standard latent models available in 
\texttt{R-INLA} and we are aware that the corresponding model could be estimated through 
the well-known Kalman Filter \citep{kalman1960new}. However, this simple model is useful to illustrate the 
framework used in this paper, which allow us to 
fit more complex dynamic models that would otherwise not be available through 
\texttt{R-INLA}. The presented approach involves an augmented model structure in
which the system equations are treated as observation equations.

The key step is to equate to zero the system equations
of the state-space model, so that
\begin{equation}
0=a_{t}-a_{t-1}-w_{t},\quad w_{t}\sim N(0,\theta_{2}),\ t=2,...,n_d.\label{eq:fakedsystemeq}
\end{equation}
Then it is possible to build an augmented model by merging the "faked
zero observations" from Eq.~\eqref{eq:fakedsystemeq} to the actual 
observations $\{y_{t},\ t=1,...,n_d\}$ of Eq.~\eqref{eq:fodlm:obseq}. 
In addition, the "faked observations" are assumed to follow a Gaussian 
distribution with high and fixed precision to represent the
fact that those artificial observations are deterministically known.
Instead of using this Gaussian distribution with high and fixed precision
and mean given by $\phi_{t}=a_{t}-a_{t-1}-w_{t}$ for the artificial
observations, as in \citet{ruiz2011direct}, we use in what follows
a Gaussian with variance $\theta_{2}$ and mean $\eta_{t}^{*}=a_{t}-a_{t-1}$.
This is an equivalent representation and will make it easier to describe 
in Section \ref{sec:Contribution} the extension of this approach to dynamic
models with non-Gaussian error terms in the system equations. 

To complete the model definition, note that
there is no information about $a_{t}$ beyond the temporal evolution
given by Eq.~\eqref{eq:fodlm:sustemeq}, and so we only need to
know the perturbations $w_{t}$, $t=2,...,n_d$ to estimate the states
$a_{t}$, since $\{w_{t}\}$ are the only stochastic term in system
equation (Eq.~\eqref{eq:fodlm:sustemeq}).
This characteristic of dynamic models allow to represent the 
dependence structure as a function of the independent perturbation 
terms. To represent this within \texttt{R-INLA}, 
let $\bs{a}=\{a_{1},...,a_{n_d}\}$ be formed by independent
random variables each following a Gaussian distribution with fixed
and low precision and encode the temporal evolution present in Eq.~\eqref{eq:fakedsystemeq}
using the copy feature available in \texttt{R-INLA} \citep{martins2012bayesian}. Finally, 
inverse-gamma priors are assigned to the variances $\theta_{1}$ and $\theta_{2}$.
The reason to use this augmented model is that it allows us
to encode the dynamic evolution of Eq.~\eqref{eq:fodlm:sustemeq}
using standard generic tools available in \texttt{R-INLA}, instead of
requiring the implementation of a different dynamic structure for 
each possible type of dynamic model.

\subsection{\texttt{R-INLA} for non-Gaussian DLM}\label{sec:Contribution}

We now present how to perform fast Bayesian inference on non-Gaussian DLM through 
the \texttt{R-INLA} package. We first formalize the augmented model described in 
Section~\ref{sec:INLAforDLMs} and the likelihood correction described in 
Section~\ref{sec:INLAreview} in this framework. We then show how our approach can 
be exploited to fit non-Gaussian DLM using \texttt{R-INLA}. The results of formalizing 
our approach overcomes the limitation assumption of independence for the non-Gaussian 
components in the latent field and, moreover, generalizes 
the DLM class of models.

The augmented model approach described in Section~\ref{sec:INLAforDLMs} 
can be represented using a hierarchical framework. Similar to Section~\ref{sec:INLAreview}, assume we 
have an augmented response vector $\bs{z}$ with $z_{t}=y_{t}$ if $t\leq n_d$ and $z_{t}=0$ if $n<t\leq2n_d-1$ and 

\begin{list}{\labelitemi}{\leftmargin=5em}
\item {\textbf{Stage 1.}} $\bs{z}|\bs{x},\bs{\theta}\sim\pi(\bs{z}|\bs{x},\bs{\theta})=\prodd{t=1}{2n_d-1}\pi(z_{t}|x_{t},\bs{\theta})$,
where \end{list} 
\begin{equation}
\pi(z_{t}|x_{t},\bs{\theta})=\Bigg\{\begin{array}{cc}
\pi(y_{t}|x_{t},\theta_{1}) & \mbox{ for }1\leq t\leq n_d\\
\pi(z_{t}|x_{t},\theta_{2}) & \mbox{ for }n<t\leq2n_d-1
\end{array}\label{eq:fakelikelihood-2}
\end{equation}
with $\pi(y_{t}|x_{t},\theta_{1})\overset{d}{=}\mathcal{N}(y_{t};x_{t},\theta_{1})$
and $\pi(z_{t}|x_{t},\theta_{2})\overset{d}{=}\mathcal{N}(0;x_{t},\theta_{2})$. Note
that, as mentioned in Section \ref{sec:INLAforDLMs}, we have used a 
Gaussian distribution with variance given by $\theta_{2}$ as the likelihood for the
artificial zero observations. Internally, for \texttt{R-INLA}, the $(4n_d-1)$-dimensional latent field is defined as 
\begin{list}{\labelitemi}{\leftmargin=5em}
\item {\textbf{Stage 2.}} $\bs{x}=(\eta_{1},...,\eta_{n_d},\eta_{n_d+1},...,\eta_{2n_d-1},a_{1},...,a_{n_d})$,
\end{list} where $\bs{a}$ is given independent Gaussian priors with
low and fixed precision, $\eta_{t}=a_{t}+s_{t}$ is the linear predictor
connected to the observation $y_{t}$, for $t=1,...,n_d$ and $\eta_{t}=a_{t}-a_{t-1}+s_{t}$
is the linear predictor connected with the artifial zero observations,
for $t=n_d+1,...,2n_d-1$, and $s_{t}$ is a small noise represented by
a Gaussian distribution with zero mean and high and fixed precision
to eliminate a rank deficiency in the above representation of $\bs{x}$.
Finally, priors are assigned to the hyperparameters of the model
\begin{list}{\labelitemi}{\leftmargin=5em}
\item {\textbf{Stage 3.}} $\theta_{1}\sim IG(a_{v},b_{v}),\quad \theta_{2}\sim IG(a_{w},b_{w}).$
\end{list}

By comparing this hierarchical representation with the likelihood correction 
approach described in Section \ref{sec:INLAreview}, we note that we are approximating
the distribution of the state vector $\bs{a} = \{a_t,\ t=1,...,n_d\}$, defined by 
Eq. (\ref{eq:fodlm:sustemeq}), which is originally given
by a Gaussian with precision matrix $Q_{a}=\theta_{2}^{-1}\bs{R}$, with

\begin{tiny} 
\begin{equation*}
\bs{R} = \left(\begin{array}{ccccccc}
1 & -1 & 0 & \cdots & 0 & 0 & 0\\
-1 & 2 & -1 & \cdots & 0 & 0 & 0\\
0 & -1 & 2 & \cdots & 0 & 0 & 0\\
\vdots & \vdots & \vdots & \ddots & \vdots & \vdots & \vdots\\
0 & 0 & 0 & \cdots & 2 & -1 & 0\\
0 & 0 & 0 & \cdots & -1 & 2 & -1\\
0 & 0 & 0 & \cdots & 0 & -1 & 1
\end{array}\right),
\end{equation*}
\end{tiny}
by a very low precision independent Gaussian distribution. In the correction approach
\[
\pi(\bs{a})\propto\text{constant}
\]
and this approximation is corrected in the likelihood function by adding
the following correction term 
\[
CT=\prod_{t=n_d+1}^{2n_d-1}\pi(z_{t}|x_{t},\theta_{2})
\]
with $\pi(z_{t}|x_{t},\theta_{2})$ defined in Eq.~\eqref{eq:fakelikelihood-2}. Note that 
this representation also corresponds to those "faked zero observations" of 
Eq.~\eqref{eq:fakedsystemeq}.
Once we have identified this, observe that the log likelihood $g_{t}(x_{t})$
and the log correction term $h_{t}(x_{t})$ in Eq.~\eqref{eq:INLAext:fullcondx}
both have quadratic forms, which implies that the full conditional of the 
latent field $\pi(\bs{x}|\bs{y},\bs{\theta})$ is Gaussian distributed, meaning 
that \texttt{R-INLA} gives exact results up to a small integration error, 
as mentioned in Section \ref{sec:INLAreview}.

Next, assume the following non-Gaussian DLM,
\begin{align}
y_{t} & =a_{t}+v_{t},\quad v_{t}\sim N(0,\theta_{1}),\ t=1,...,n_d\label{eq:nGfodlm:obseq}\\
a_{t} & =a_{t-1}+w_{t},\quad w_{t}\sim t(0,\tau,\nu),\ t=2,...,n_d,\label{eq:nGfodlm:sustemeq}
\end{align}
which can be written in a hierarchical structure
\begin{align*}
y_{t}\mid a_{t},\theta_{1} & \sim N\left(a_{t},\theta_{1}\right)\\
a_{t}\mid a_{t-1},\tau,\nu & \sim t\left(a_{t-1},\tau,\nu\right)\\
\tau\sim\pi\left(\tau\right),\quad\nu & \sim\pi\left(\nu\right),\quad \theta_{1}\sim\pi\left(\theta_{1}\right)
\end{align*}
highlighting the fact that the latent field is no longer Gaussian.
Note that the $\bs{\theta_{2}}=(\tau,\nu)$ and the distribution of $v_{t}$ in Eq.~\eqref{eq:nGfodlm:obseq}
could have a non-Gaussian distribution as well, since non-Gaussian
likelihood functions are already standard in \texttt{R-INLA}, but
using a Gaussian here makes the final impact of the non-Gaussianity
of $w_{t}$ more easily visible and analyzed in Section~\ref{sec:Simulations}. 
As mentioned in Section~\ref{sec:Introduction}, the motivation of using 
heavier tailed distributions such as Student-t in the noise
of the latent system is to robustify the model. Robustifying the
model means that the dynamic system is less sensitive to different types of 
outliers. By allowing this higher flexibility of $w_{t}$ we can better  
handle what is called \textit{innovative outliers} in time series literature 
\citep{fox1972,masreliezmartin1977,mcquarrietsai2003}.

By a similar argument made in Section~\ref{sec:INLAforDLMs}, we note that we
only need to know the stochastic terms $\{w_t,\ t=2,...,n_d\}$ and the
system dynamics in Eq.~\eqref{eq:nGfodlm:sustemeq} to estimate $\bs{a}=(a_{1},...,a_{n_d})$.
Consequently, if we include those pieces of information in the likelihood
function through a correction term, we can assign independent Gaussian
priors with zero mean and low and fixed precisions for $\bs{a}$,
which will lead to the following hierarchical model 
\begin{list}{\labelitemi}{\leftmargin=5em}
\item {\textbf{Stage 1.}} $\bs{z}|\bs{x},\bs{\theta}\sim\pi(\bs{z}|\bs{x},\bs{\theta})=\prodd{t=1}{2n_d-1}\pi(z_{t}|x_{t},\bs{\theta})$,
where \end{list} 
\begin{equation*}
\pi(z_{t}|x_{t},\bs{\theta})=\Bigg\{\begin{array}{cc}
\pi(y_{t}|x_{t},\theta_{1}) & \mbox{ for }1\leq t\leq n_d\\
\pi(z_{t}|x_{t},\tau,\nu) & \mbox{ for }n_d<t\leq2n_d-1
\end{array}
\end{equation*}
with $\pi(y_{t}|x_{t},\theta_{1})\overset{d}{=}\mathcal{N}(y_{t};x_{t},\theta_{1})$
and $\pi(z_{t}|x_{t},\tau,\nu)\overset{d}{=}t(0;x_{t},\tau,\nu)$.
\begin{list}{\labelitemi}{\leftmargin=5em}

\item {\textbf{Stage 2.}} $\bs{x}=(\eta_{1},...,\eta_{n_d},\eta_{n_d+1},...,\eta_{2_dn-1},a_{1},...,a_{n_d})$
\end{list} where $\bs{a}$ and $\bs{\eta}$ are the same as defined earlier. Finally, priors are assigned to the hyperparameters of the model \begin{list}{\labelitemi}{\leftmargin=5em}

\item {\textbf{Stage 3.}} $\bs{\theta}\sim\pi(\bs{\theta})$ with
$\bs{\theta}=(\theta_{1},\tau,\nu)$ \end{list}

We see that the hierarchical model above is very similar to the one presented
by Eq. (\ref{eq:fakelikelihood-2}) and the difference is on the correction
term 
$$CT=\prod_{t=n_d+1}^{2n_d-1}\pi(z_{t}|x_{t},\tau,\nu),$$
which is no longer Gaussian distributed, leading to a full conditional
of the form Eq.~\eqref{eq:INLAext:fullcondx} with a non-Gaussian log
correction term $h_{t}(x_{t})$. As we have showed, this configuration fits
the framework summarized in Section \ref{sec:INLAreview} and therefore we can apply 
the results 
to the context of 
non-Gaussian dependent latent fields, specifically DLMs. Thus, \texttt{R-INLA} provides 
accurate results for non-Gaussian DLM, as long as the non-Gaussian distribution 
attributed to the error terms of the system equations are not too far from a 
Gaussian distribution, as discussed in Section~\ref{sec:Introduction}. 
This assumption is satisfied by the Student-t distribution, 
as well as for other distributions that corrects the Gaussian in terms of
skewness and/or kurtosis.

\section{Simulation Study}\label{sec:Simulations}

In this Section we present the results of a Monte Carlo simulation for the Toy Example 
defined in Section~\ref{sec:Contribution} (see Eqs.~\eqref{eq:nGfodlm:obseq} and 
\eqref{eq:nGfodlm:sustemeq}) to better understand the benefits of fitting a non-Gaussian 
DLM with INLA. Moreover, we investigate the property of different model selection criteria
available from \texttt{R-INLA} in this context. 
We have chosen to perform a contamination study similar to the ones presented in \citet{pinheiro2001efficient} 
and in \citet{martins2014extendingScan} where the noise $w_{t}$ from Eq.~\eqref{eq:nGfodlm:sustemeq}
is contaminated with the following mixture of Gaussian distributions 
\begin{equation*}
w_{t}\sim\left(1-p\right)\times N\left(0,\theta_{2}\right)+p\times f\times N\left(0,\theta_{2}\right)\quad t=1,\ldots,n_d,
\end{equation*}
where $p$ is the expected percentage of \textit{innovative outliers}
in the latent system and $f$ is a fixed value indicating the magnitude
of the contamination. We have generated all possible scenarios with 
$n_d=100,250,500$, $p=0,0.05,0.1,0.15,0.20,0.25$ and $f=2,4,8$, resulting in a total 
of $54$ different scenarios. For each of them, $1000$ datasets were simulated and analyzed.
The true variance parameter of the observational and system noises are set to $\theta_{1},\theta_{2}=2$. 

In \texttt{R-INLA} the Student's t likelihood is parametrized in terms of its marginal 
precision $\tau$ and degrees of freedom $\nu$. This is advantageous because the precision 
parameter under the Gaussian and the Student's t distribution possess the same interpretation 
allowing the same prior to be used for $\tau$ whether we refer to the Gaussian or to the Student's t model.
In this Monte Carlo experiment we have used a 
Gamma\footnote{if $X\sim Gamma\left(a,b\right)$ then $E\left(X\right)=\frac{a}{b}$} 
prior with shape and rate parameters given by 1 and 2.375 for both the observational and system 
noise precision parameters. The prior for $\nu$ is based on the framework of 
\cite{martins2013priorarXiv}. In their context, the prior is design 
for the flexibility parameters, which in this case is the degrees of freedom $\nu$, in
such way that the basic model plays a central role in the more flexible one. In our 
context, it means that the prior for the degrees of freedom is constructed such that 
the mode of the prior happens to be in the value that recovers the Gaussian model and 
deviations from the Gaussian model are penalyzed based on the distance between the 
basic and the flexible model. The prior specification consists in the choice of the degree of 
flexibility (df) parameter, $0<\text{df}<1$, which represents the percentage of prior mass 
attributed to the degrees of freedom between 2 and 10. We have set $\text{df}=0.3$ in our 
applications. We refer to \cite{martins2013priorarXiv} for more details about priors for flexibility parameters.

As mentioned in the introduction, a model based on the Student's t distribution is expected 
to be more robust with respect to outliers in a contaminated data setup when compared to a 
similar model based on Gaussian distributions. To assess the gain in performance of the more 
flexible model based on the Student's t distribution, we will compute the mean squared 
error (MSE), the conditional predictive ordinate (CPO) \citep{gelfanddey1992,deychan1997} 
and deviance information criteria (DIC) \citep{Spiegelhalter2002}.
The intuition behind the CPO criterion is to choose a model with higher predictive power 
measured in terms of predictive density.


For the $j$-th simulated dataset of a given scenario, let $a_{tj}$ be the true 
latent variable at time $t$. We will denote by $\widehat{a_{tj,G}}$ and 
$\widehat{a_{tj,T}}$ the posterior mean of $a_{tj}$ computed by the Gaussian
and Student's t model, respectively. The Student-t model efficiency over the 
Gaussian one to estimate $a_{tj}$ for each dataset $j$ is defined by
\begin{equation*}
E_{j}=\frac{\sum_{t=1}^{n_d}\left(\widehat{a_{tj,G}}-a_{tj}\right)^{2}}{\sum_{t=1}^{n_d}\left(\widehat{a_{tj,T}}-a_{tj}\right)^{2}} - 1,
\end{equation*}
which can be viewed as ratio of the respective MSEs centered at $0$.

\begin{figure}[htb]
 \begin{center}
 \includegraphics[scale=0.45]{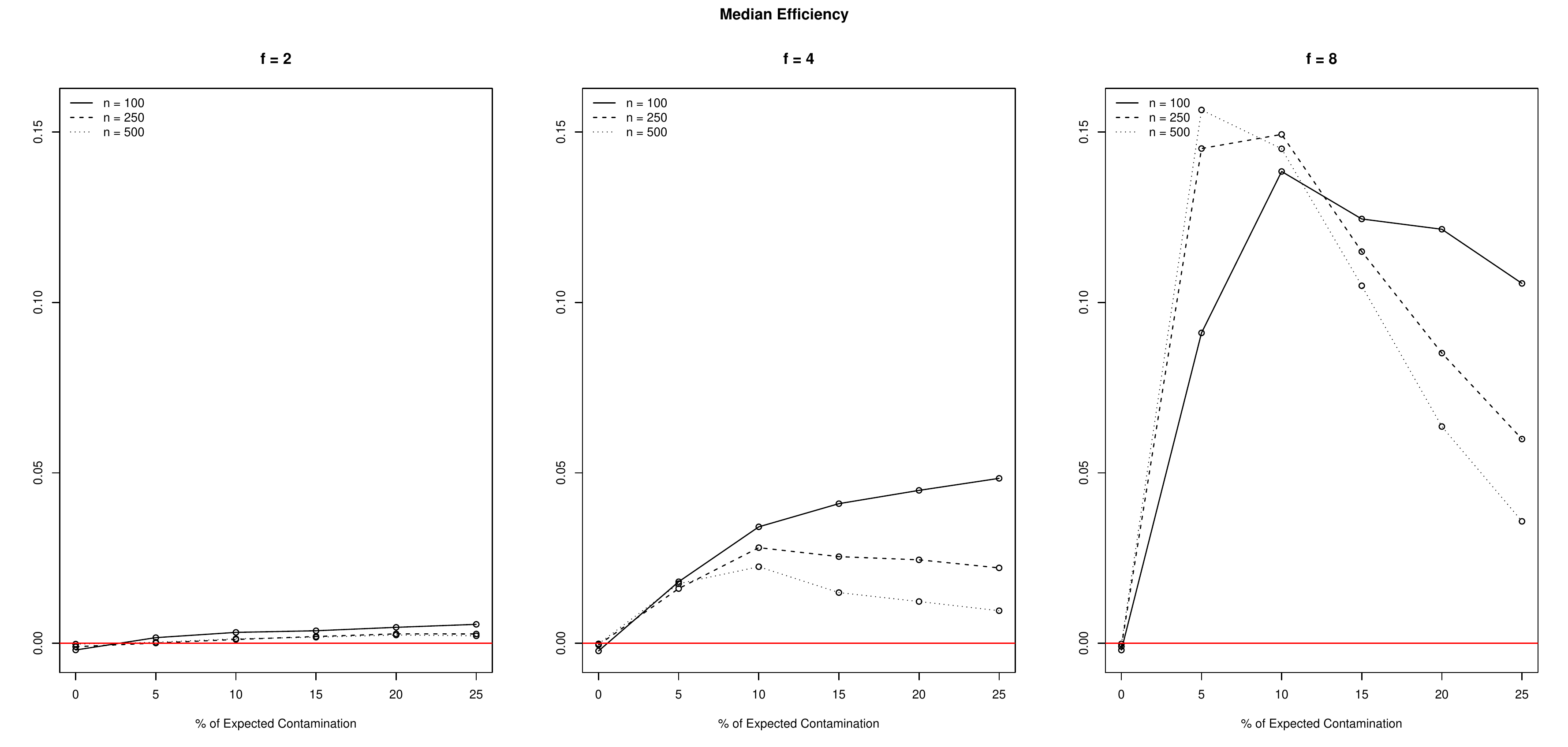}
 \end{center}
 \caption{Median of Efficiencies for magnitude f = 2 (left), f = 4 (center) and f = 8 (right), 
          $n_d$ = 100 (solid line), $n_d$ = 250 (dashed line) and $n_d$ = 500 (dotted line). We have the percentage
          of expected contamination in the x-axis and the median of efficiency in the y-axis.}
 \label{fig:MSEmedians}
 \end{figure}

Figure \ref{fig:MSEmedians} represents the median over $\{E_j,\ j = 1, ..., 1000\}$ for each scenario.
The results were as expected. There were slight efficiency improvements for close contamination patterns
while the efficiency gains become larger as we move to higher contamination patterns, reaching efficiency 
gains greater than $15\%$ for some critical scenarios. The efficiency gains are higher for moderate 
expected contamination percentage, around $10\%$ in our case, and this non-monotonic behavior can be 
explained by the fact that once the data becomes too much contaminated, not even the more flexible
model based on the Student's t distribution can continue to give increasingly better results when compared 
to the Gaussian model, although the more flexible one continues to improve upon it.

\begin{figure}
\begin{center}
\includegraphics[scale=0.45]{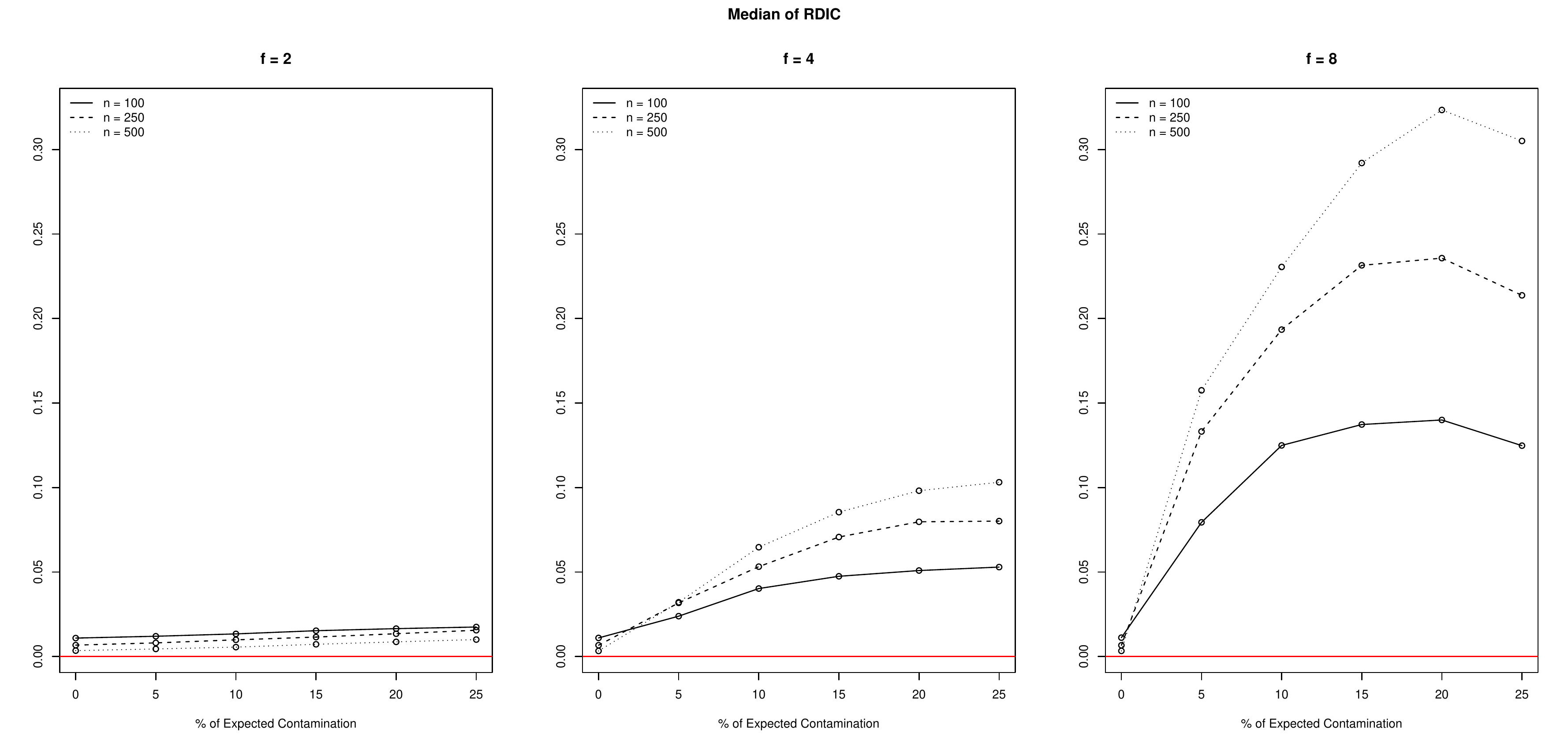}
\includegraphics[scale=0.45]{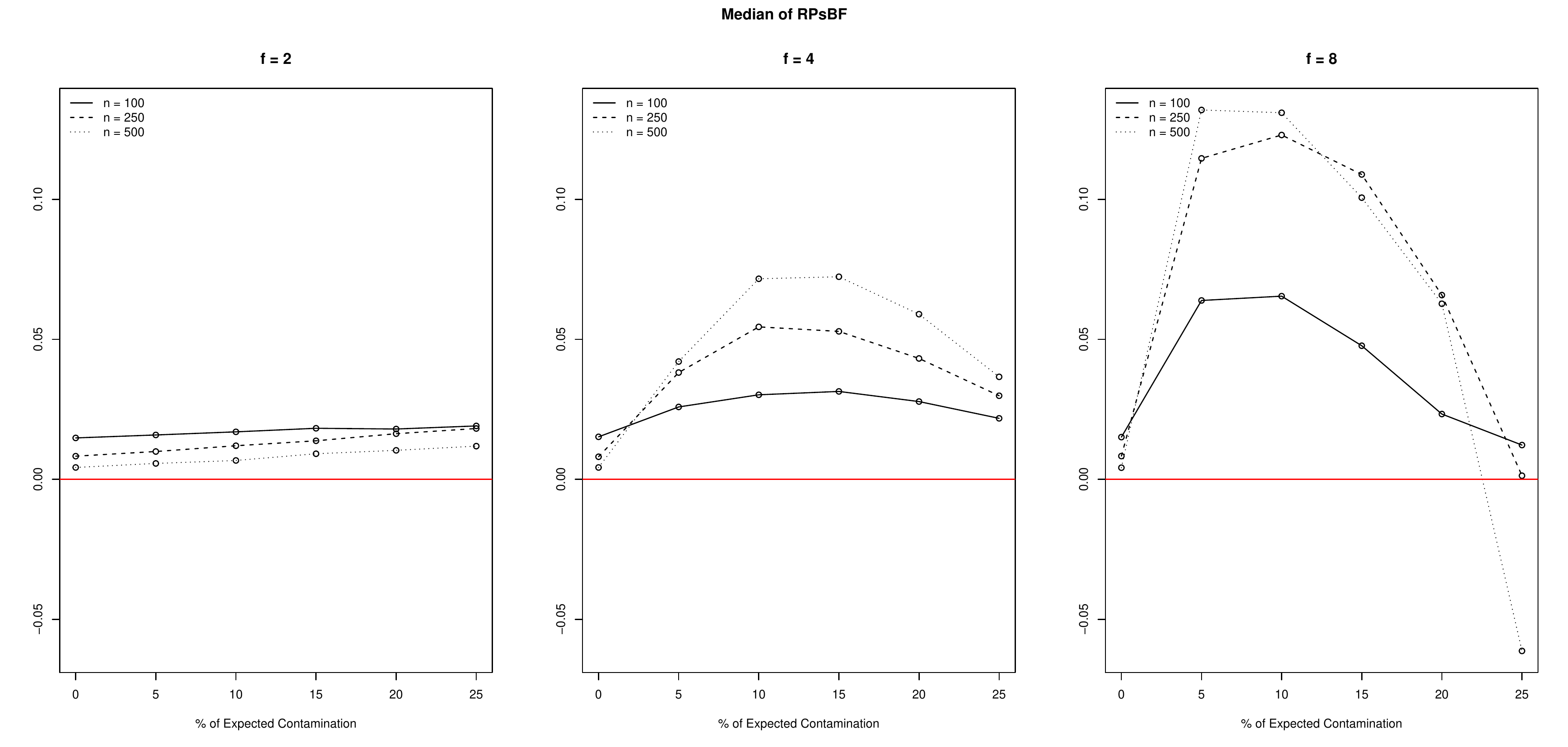}
\end{center}
\caption{Top: Median of RDIC, in the y-axis, for all scenarios; Bottom: Median of RPsBF, in the 
y-axis, for all scenarios. We have the percentage of expected contamination in the x-axis and 
all scenarios are: f = 2 (left), f = 4 (center) and f = 8 (right), $n_d$ = 100 (solid line), $n_d$ = 250 
(dashed line) and $n_d$ = 500 (dotted line).}
\label{fig:criteria}
\end{figure}

To compare the model fitting we use the DIC as well as the CPO criteria. First, we define the
relative DIC (RDIC) as
\begin{equation}
\mbox{RDIC}_{j}=\frac{\mbox{DIC}_{Gj}-\mbox{DIC}_{tj}}{\mbox{DIC}_{tj}},
\label{eq:RDIC}
\end{equation}
for each one of the simulated data, $j=1,\ldots,1000$.
In the top part of Figure~\ref{fig:criteria} we plot the median of RDIC values obtained by the 
fitted Gaussian model and by fitted the Student's t model for each scenario. From this Figure 
we observe the same pattern of Figure~\ref{fig:MSEmedians}. 

The summary statistic provided by the CPO criteria is called logarithm of the pseudo marginal likelihood (LPML) which evaluates the predictive power of a model. Therefore, to compare both models the LPML difference is used. To make it comparable to other goodness-of-fit measures, e.g. DIC, we define the -LPML by
\begin{equation*}
\mbox{-LPML}_{j}=-\left(\sum_{i=1}^{n_d}\log\left\{ \pi\left(y_{i}\mid\boldsymbol{y}_{-i}\right)\right\} \right)_{j}
\end{equation*}
where $j$ is the $j$-th dataset in a given scenario. In this definition, lower values of -LPML indicates better predictive power. In order to compare both approaches, we have computed the logarithm of the 
Pseudo Bayes Factor (lPsBF) \citep{geiseddy1979} for each iteration. This measure is defined as
\begin{equation*}
\mbox{lPsBF}_{j}= \mbox{-LPML}_{tj} -(\mbox{-LPML}_{Gj}) = \mbox{LPML}_{Gj} -\mbox{LPML}_{tj}.
\end{equation*}
To make the comparison equivalent to the RDIC presented in Eq.~\eqref{eq:RDIC} we define the relative 
lPsBF (RPsBF) as
\begin{equation*}
\mbox{RPsBF}_{j}=\frac{\mbox{LPML}_{Gj}-\mbox{LPML}_{tj}}{\mbox{LPML}_{tj}}.
\end{equation*}

From the bottom part of Figure~\ref{fig:criteria}, all conclusions from the MSE and RDIC 
can be applied in the context of the RPsBF measure, but the 
gain becomes more evident. Moreover, we can see from the bottom part of Figure~\ref{fig:criteria} that
when the simulated scenario is stable with low expected proportion and low contamination,  
the median of the RPsBF is small and not significant. However, for larger sample size and 
contamination it is showed that the Student-t approach is preferable for most of the scenarios and 
highlights this choice when the magnitude of the contamination increase reaching values of this median 
relative difference even higher than 10\% in some cases. One curious fact observed is that in the most critical 
scenario where $p=0.25$, $f=8$ and $n_d=500$ the RPsBF values pointed incisively to the Gaussian approach, 
indicating that, since the generation process has too much contamination and generates to many 
innovative outliers, even the Student-t approach is not able to control for this behavior producing 
predictive measures that are less accurate. 

From the simulation study we can conclude that the more flexible model is preferred over the traditional one 
in most of the scenarios analyzed, and the gap between the models are higher when a moderate number of innovative 
outliers are involved. 

\section{Data Application}
\label{sec:data_application}

The goal of this Section is to analyze Brazilian Homicide rates 
with approximate Bayesian Inference for Dynamic Models using \texttt{R-INLA}. 
The data comprises homicide rates of all the Brazilian cities available online 
in the database of the Brazilian National Public Health System\footnote{www.datasus.gov.br/} (DATASUS). 
The time series under study represents the number of events standardized by each city population. 
In Brazil, experimental studies capable of evaluating the impact of determined political program 
about the spatial or temporal crime pattern are not common, mainly because of the difficulty in 
getting such data. On the other hand, international experience points-out that crime prevention 
policies are efficient after considering characteristics in which crime 
occurs \citep{sherman1995,sherman1997,sherman1998}.

According to \citet{Waiselfisz2012} and due to Brazilian legislation number 6,216 of 1975, all death, 
due to natural causes or not, must have a corresponding death register. This death register, made
by a legist or two witnesses, has a standard national structure containing
age, sex, civil state, occupation, naturality and local residence
of the victim. According to the law, the death register must be computed
by the "local of the death", i.e., the local of the occurrence
of the event. All death occurrences in the dataset have been classified by the Brazilian government 
as homicides and comprises annual time series from 1980 to 2010.

Homicides studies have a broad field of sociological research (see,
for example, \citet{jacobs2008}). Homicides represent a specific criminal 
category that, although it has less cases when compared to patrimonial crimes, 
as burglary and robbery, it generates strong population demand for public policies of 
prevention. In this sense, studies that deal with the temporal dynamic 
of homicides try to associate, in a general way, this behavior with economic, 
social and political factors. For instance, high Brazilian homicide rates could be 
due to high levels of unemployment, poverty and economic inequality \citep{mir2004}. 
Population age structure \citep{graham1995,floodpage2000} and disordered 
populational growth, or inequality in social conditions, could be other factors to 
explain these rates \citep{blau1982}.
In summary, alteration in the criminal historic behavior 
is associated to \textit{law-enforcement} elements, such as increase of the number 
of police officers, expenses with safety policies and increase of imprisonment rates. 
Analyzing data from New York City, \citet{zimring2007} concluded 
that "there is a strong evidence that changing the number of cops, as well 
policing tactics, has a important impact in crime" (pg. 151). Specifically in 
Brazil, \citet{goertzel2009}, while studying the behavior of the strong decline 
of homicides rates in S\~{a}o Paulo state since 2000's, have concluded that more 
repressive police models and disarmament policies reduced substantially homicides and other 
violent crimes in the state. Statistically, it was expected that in certain 
regions, the rates could suffer from sudden structural changes. This fact 
requires a robust approach to model them, such as the assumption of heavy-tailed 
distribution for the latent system noise to handle possible innovative outliers as 
discussed in Section~\ref{sec:Simulations}.

The model adopted was similar to Eqs.~\eqref{eq:fodlm:obseq} and \eqref{eq:fodlm:sustemeq}.
However, to model each time series we considered the capital cities grouped according 
to a specific criterion. Specifically, we have used the following capital division:
\begin{itemize}
\item Group 1 (G1) - S\~{a}o Paulo and Rio de Janeiro
\item Group 2 (G2) - Belo Horizonte, Recife, Vit\'{o}ria and Porto Alegre
\item Group 3 (G3) - All the 21 remaining capitals.
\end{itemize}
This division was motivated in terms of different urbanization process. For each group 
created we have one tendency estimated for the capitals and one tendency estimated for 
all first order spatial neighbors (those which share border with the capital). Thus, 
for each group, there is a model for the capitals and another one for the capitals 
neighbors, totaling six models.

Let $Y_{til}$ be the Homicide Rate of city $i=1,\ldots,n_l$ in group $l=1,\ldots,6$ at time $t$, 
then we have the following model:
\begin{align*}
Y_{til} & =a_{tl}+v_{til}, & \, v_{til}\sim N(0,\theta_{1il}), & \, & t=1,...,31\\
a_{tl} & =a_{(t-1)l}+w_{tl}, & \, w_{tl}\sim(0,\theta_{2l}), & \, & t=2,...,31 
\end{align*}
where $w_{tl}$ is either Gaussian or Student-t and $n_l$ is the number of cities in group $l$.

To complete the model specification it is necessary to specify the priors for 
the hyperparameters. In our case, we have to specify priors for the precision 
of each group in the observational equation and a single prior for the system equation precision. 
The prior of the precision of the observational equation is created to cover with 
high probability the variances of all Brazilian cities, setting a prior 
$\theta_{1il}^{-1}\sim Gamma\left(5,500\right), i = 1,\ldots,n_{l}$ and $l = 1, \ldots, 6$
we cover with 90\% of probability the values between the 25th and 75th quantiles 
of the cities sample variance. For the precision prior for the latent equation, a prior 
$\theta_{2l}^{-1}\sim Gamma\left(1,0.1\right)$ is used for all cases. Finally, the same prior 
set in the simulation study for the degrees of freedom $\nu$ assuming 30\% of 
probability of prior mass for $\nu$ values between 2 and 10 is used.

It is important to emphasize that the latent state represented by the vector 
$\boldsymbol{a}$, can be interpreted as the non-observed mean tendency of the 
cities of each modeling.

\begin{figure}[H] 
\begin{center}
\includegraphics[scale=0.35]{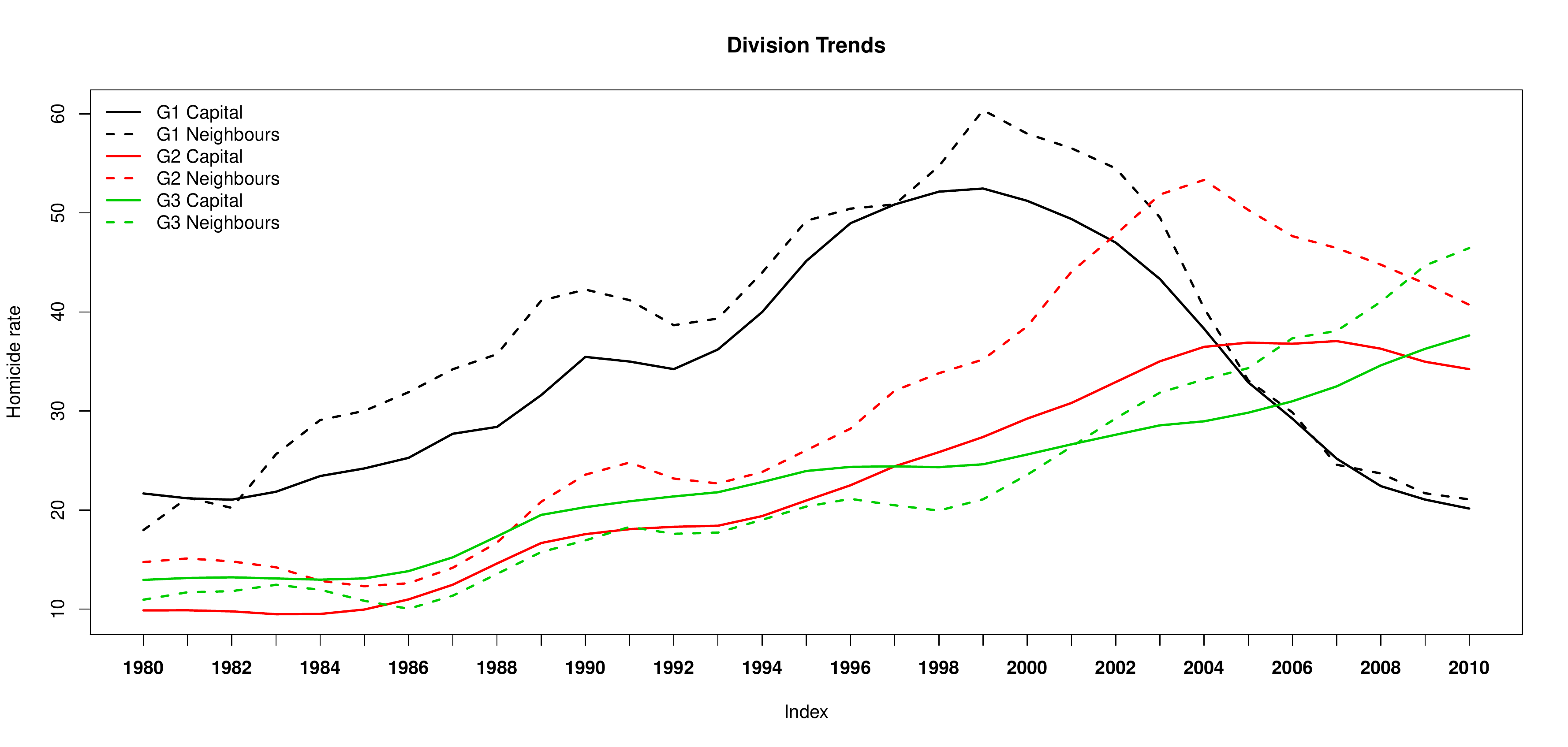}
\par\end{center}
\caption{Division posterior tendencies, $\boldsymbol{a}$, for each group of Student-t approach. G1: S\~{a}o Paulo and 
Rio de Janeiro; G2: Belo Horizonte, Recife, Vit\'{o}ria and Porto Alegre; G3: All the 21 remaining capitals}
\label{fig:group}
\end{figure}

To understand the temporal trends, sociological literature analyzes 
the context of how interpersonal violence spreads. Analyzing a historical 
time series of more than 30 years of delinquency and crime, \citet{shaw1942} 
verified that not only crime, but several social problems were related to a 
disorganized social environment. The theoretical approach developed by these 
authors helps to understand the effects of a unplanned urbanization process 
in criminal behavior. In Brazil, specially in big cities, the economic development 
was followed by the appearance of urban enclaves (such as slums) where the impairment 
of the traditional mechanisms of social control promotes an environment of 
differentiated criminal opportunities \citep{sutherland1992}. For more information 
about \textit{Social Disorganization Theory} we refer to \cite{shaw1942} and \cite{kubrin2003}.

Figure~\ref{fig:group} presents the posterior tendency, $\boldsymbol{a}$, of each group capital and neighbours for 
the adjusted Student-t model. The tendency of the first group tells us that its urbanization 
process started earlier when compared to the other groups because the homicide rate started to 
get higher firstly. Another aspect was that until 2000, all the tendencies were nearly linear for 
all groups. However, a reverse tendency was observed as result of investments and criminal control 
policies established in G1 \citep{goertzel2009}. The other two groups are heading towards the same 
behavior, but they still didn't show it in such an evident way, as seen in G1, the effect of safety 
policies. In G2 some states already adopted some safety measures, e.g., Minas Gerais, which  
capital is Belo Horizonte, with the creation of \textit{Integra\c{c}\~{a}o da Gest\~{a}o em Seguran\c{c}a P\'{u}blica} 
(IGESP\footnote{www.sids.mg.gov.br/igesp}) in May 2005, and Pernambuco, which capital is Recife, 
with the creation of the \textit{Pacto Pela Vida}\footnote{www.pactopelavida.pe.gov.br/pacto-pela-vida} 
in May 2007. 

\begin{table}[H]
\begin{centering}
\begin{tabular}{c|c|cc|cc|cc}
\cline{3-8} 
\multicolumn{1}{c}{} & \multicolumn{1}{c}{} & \multicolumn{2}{c}{G1} & \multicolumn{2}{c}{G2} & \multicolumn{2}{c}{G3}\tabularnewline
\cline{3-8} 
\multicolumn{1}{c}{} & \multicolumn{1}{c}{} & Gaussian & Student-t & Gaussian & Student-t & Gaussian & Student-t\tabularnewline
\hline 
-LPML & Capitals & 229.09 & 227.81 & 496.72 & 496.45 & 2481.70 & 2473.55\tabularnewline
\cline{2-8} 
 & Neighbours & 3536.78 & 3517.38 & 2379.98 & 2375.89 & 3026.54 & 3002.48\tabularnewline
\hline 
DIC & Capitals & 460.22 & 459.64 & 997.44 & 996.60 & 4960.98 & 4946.32\tabularnewline
\cline{2-8} 
 & Neighbours & 7063.15 & 7030.68 & 4753.67 & 4746.94 & 6040.70 & 6004.27\tabularnewline
\hline 
\end{tabular}
\par\end{centering}
\caption{Quality measures of the division. G1: S\~{a}o Paulo and Rio de Janeiro;
G2: Belo Horizonte, Recife, Vit\'{o}ria and Porto Alegre; G3: All the
21 remaining capitals}
\label{tab:Measures1division}
\end{table}

In order to assess the goodness of fit of our robust approach, we computed the -LPMLs and the DICs for every model,  
as can be seen in Table~\ref{tab:Measures1division}. From Table~\ref{tab:Measures1division}, 
all criteria, -LPMLs and DICs, pointed to the robust approach assuming the Student-t distribution 
for the system noise. To verify the evidence that the robust approach outperforms significantly the traditional one, 
in the real data set, we chose to analyze Table~\ref{tab:PsBF} which was 
proposed and used in \citet{prates2010}. From Table~\ref{tab:PsBF}, we can see that the PsBF, which 
is the -LPML difference as presented in Section~\ref{sec:Simulations}, have a positive 
evidence against the Gaussian model for the capital modeling in G1 (2PsBF = 2.56) and a strong evidence 
against the Gaussian model for the neighboring modeling in G2 (2PsBF = 8.18). However, the 
bigger gain in favor of the Student-t model is verified for the neighboring modeling in G1 and G3 as 
well as the capitals in this last group, with all having strong evidence in terms of the -LPML 
differences. Since for each group we have many first order neighbors time series, for G1, G2 and G3 
there are 30, 24 and 152 neighbors respectively, analysis suggests that as the number of cities 
increases there is a demand for a more robust approach.

\begin{table}[H]
\begin{center}
\begin{small}
\begin{tabular}{cc}
\hline 
2PsBF & Evidence against Gaussian\\
\hline 
(-1,1{]} & worth mention\tabularnewline
(1,5{]} & positive\tabularnewline
(5,9{]} & strong\tabularnewline
(9,$\infty$) & very strong\tabularnewline
\hline 
\end{tabular}
\end{small}
\end{center}
\caption{Pseudo Bayes Factor criteria}
\label{tab:PsBF}
\end{table}

In our real data application the gain in terms of predictive power 
was very clear. We also should point out that there is evidence of deviation 
from Gaussianity when we look to the posterior distribution of the $\nu$ in the 
Student-t model. From Table~\ref{posterior_dof} we can see the posterior median and 95\% 
credible intervals (CI). The median measure indicates that the Student-t distribution 
is concentrated in medium values of the degrees of freedom but the 95\% CI are 
highly asymmetric reaching very high values.

\begin{table}[H]
\begin{centering}
\begin{tabular}{c|c|c|c}
\hline 
\multicolumn{1}{c}{Group} & \multicolumn{1}{c}{Type} & \multicolumn{1}{c}{Median} & 95\% Credible Interval\tabularnewline
\hline 
G1 & Capitals & 35.53 & (5.84; 441.30)\tabularnewline
\cline{2-4} 
 & Neighbours & 28.46 & (6.04; 339.38)\tabularnewline
\hline 
G2 & Capitals & 39.93 & (6.73; 533.77)\tabularnewline
\cline{2-4} 
 & Neighbours & 31.69 & (5.56; 445.12)\tabularnewline
\hline 
G3 & Capitals & 32.23 & (5.14; 434.54)\tabularnewline
\cline{2-4} 
 & Neighbours & 55.91 & (21.69; 556.67)\tabularnewline
\hline 
\end{tabular}
\par\end{centering}
\caption{Posterior measures of the $\nu$ of Student-t approach.
G1: S\~{a}o Paulo and Rio de Janeiro; G2: Belo Horizonte, Recife, Vit\'{o}ria
and Porto Alegre; G3: All the 21 remaining capitals}
\label{posterior_dof}
\end{table}

\section{Conclusions}

This paper describes how to perform Bayesian inference using \texttt{R-INLA}
to estimate non-Gaussian Dynamic Models when the evolution noise has a non-Gaussian distribution. 
Such models can be viewed as part of latent hierarchical models where a non-Gaussian Random Field 
is assumed for the latent field and, therefore, invalidates the direct use of the INLA methodology
that requires that the latent field must be Gaussian. 

Using a random walk example we presented how to use an augmented structure to overcome the 
Gaussian limitation of INLA for the latent field.  
The key to understand why our approach works relies on the fact that we approximate the non-Gaussian 
latent field through a Gaussian distribution and corrects this approximation in the likelihood 
function trying to minimize the loss of this approximation for dependent models. We discussed 
and explained the reasons to make this approximative approach and, specially, where in the 
\texttt{R-INLA} calculations it will impact.

Through simulations, we showed the necessity of more robust models when the time series suffer sudden 
structural changes. From our results we observe that Gaussian models are sensitive to structural 
changes while our approach assuming a Student-t field is robust. Specifically, our simulation
study presented an incisive demand to avoid the usual Gaussian assumption in most contaminated 
scenarios. There are indication that some public policies for crime control can generate a positive 
effect in crime's temporal tendencies allowing the presence of structural changes. It is evident 
that other control factors might help to confirm this hypothesis, however it is very likely that 
investments in security policies, such as those implemented in G1 and G2, have contribution in the 
dynamic observed. Our homicide rate application pointed-out, as expected, that public policies could 
play an important role to explain homicides dynamics through a robust approach due to the 
characteristic of these kind of data. Although we analyzed 
homicide rate because of their sociological impacts, we are aware that this 
extension would also be well justified in other fields such as stochastic volatility models 
\citep[see, for example,][]{jacquier2003}.

As mentioned in Section~\ref{sec:methodology} a natural extension of the model class presented 
is the DGLM, where one could assume a non-Gaussian distribution for the observed data and, 
consequently, impacting Eq.~\eqref{eq:INLAext:fullcondx} which both $g_{t}(x_{t})$ and $h_{t}(x_{t})$ 
could have a non-quadratic form. This extension is investigated in a different 
manuscript. The main advantage of the model structure presented here is that it allows users 
to fit basically any complex structured non-Gaussian dynamic model with fast and good accuracy 
using a friendly tool already available.

We believe that the applied community can make good use of this methodology when necessary. For 
real time series data is not rare to observe structural breaks and a robust 
approach, as the one presented, may be more adequate to adjust this type of data.
Furthermore, we have formalized how to use the \texttt{R-INLA} software for non-Gaussian 
dynamic models in a simple way.

\bibliographystyle{apalike}
\bibliography{bibliografia_all}

\end{document}